\documentclass[amssymb,prb,aps,twocolumn,amsmath,showpacs]{revtex4}

\usepackage[dvips]{graphicx}

\usepackage{bm}

\begin{document}

\title{Measurement of spin memory lengths in PdNi and PdFe ferromagnetic alloys}
\author{H. Arham}
\author{T. S. Khaire}
\author{R. Loloee}
\author{W. P. Pratt, Jr.}
\author{Norman O. Birge}
\email{birge@pa.msu.edu}

\affiliation{Department of Physics and Astronomy, Michigan State
University, East Lansing, MI 48824-2320}
\date{\today}

\begin{abstract}
Weakly ferromagnetic alloys are being used by several groups in
the study of superconducting/ferromagnetic hybrid systems. Because
spin-flip and spin-orbit scattering in such alloys disrupt the
penetration of pair correlations into the ferromagnetic material,
it is desirable to have a direct measurement of the spin memory
length in such alloys. We have measured the spin memory length at
4.2\,K in sputtered Pd$_{0.88}$Ni$_{0.12}$ and
Pd$_{0.987}$Fe$_{0.013}$ alloys using methods based on
current-perpendicular-to-plane giant magnetoresistance.  The
alloys are incorporated into hybrid spin valves of various types,
and the spin memory length is determined by fits of the Valet-Fert
spin-transport equations to data of magnetoresistance vs. alloy
thickness.  For the case of PdNi alloy, the resulting values of
the spin memory length are $l_{sf}^{PdNi}=2.8\pm 0.5$\,nm and
$5.4\pm 0.6$\,nm, depending on whether or not the PdNi is exchange
biased by an adjacent Permalloy layer.  For PdFe, the spin memory
length is somewhat longer, $l_{sf}^{PdFe}=9.6\pm 2$\,nm,
consistent with earlier measurements indicating lower spin-orbit
scattering in that material. Unfortunately, even the longer spin
memory length in PdFe may not be long enough to facilitate
observation of spin-triplet superconducting correlations predicted
to occur in superconducting/ferromagnetic hybrid systems in the
presence of magnetic inhomogeneity.
\end{abstract}


\maketitle

The interplay between ferromagnetism and superconductivity has
piqued the interest of physicists for several
decades.\cite{deGennes:66}  In hybrid
superconducting/ferromagnetic (S/F) systems, the strong exchange
field in the F material limits the penetration of superconducting
pair correlations (the proximity effect) to very short distances,
of order $\xi_F = \surd{\hbar D/E_{ex}}$ in the dirty limit, where
$D$ and $E_{ex}$ are the diffusion coefficient and exchange
energy, respectively, in the F material.\cite{BuzdinReview}
Several groups studying S/F systems have used weakly ferromagnetic
alloys to reduce $E_{ex}$ and hence increase
$\xi_F$.\cite{Ryazanov:01,Kontos:01} Ryazanov and
co-workers\cite{Oboznov:06} fabricated S/F/S Josephson junctions
using CuNi alloy as the F material, and found a very rapid
decrease of the critical current with CuNi thickness, probably due
to spin-flip scattering from small magnetic clusters in the
inhomogeneous magnetic alloy.\cite{Pratt:unpub}  Kontos \textit{et
al.}\cite{Kontos:02} found a somewhat slower decrease in critical
current vs. thickness of S/I/F/S junctions with PdNi alloy as the
F material, raising the hope that the spin memory length in PdNi
might be considerably longer than in CuNi.  If that were true, it
would open the possibility to observe an unusual type of proximity
effect in which spin-triplet pair correlations penetrate deep into
the F metal, a situation predicted to occur in S/F systems in the
presence of certain forms of magnetic
inhomogeneity.\cite{Bergeret:03} Clearly, direct measurements of
the spin memory length in PdNi and other ferromagnetic alloys will
not only help in the interpretation of existing data on S/F
systems, but may also indicate which materials are most likely to
be fruitful in future experiments searching for the predicted
triplet correlations.\cite{Bergeret:03}

The spin memory length, $l_{sf}$, also plays an important role in
the context of giant magnetoresistance (GMR) of metallic
multilayers.\cite{ValetFert:93}  For example, spin memory loss in
the nonmagnetic spacer layer (N) in an F/N/F spin valve reduces
the GMR signal, whereas spin memory loss strategically placed in
layers outside the active F/N/F core can actually enhance the GMR
signal.\cite{Urazhdin:04} In ferromagnetic materials, spin memory
loss tends to limit the thickness of the F layer that actively
contributes to the GMR signal.\cite{BassPrattSection} Measurements
of $l_{sf}$ have been performed on a wide variety of N materials
and on a handful of F materials.\cite{BassPrattReview}  In this
paper, we present measurements of the spin memory length in
specific PdNi and PdFe alloys, chosen for reasons stated below.

We chose initially to study Pd$_{1-x}$Ni$_x$ alloy with $x = 12$
atomic $\%$, because alloys with concentrations close to that
value have been used previously in several studies of S/F
systems.\cite{Kontos:01,Kontos:02,Cirillo:06,Matsuda:07} The PdNi
sputtering target consists of a 5.6-cm diameter target of pure Pd
with several 6-mm diameter Ni plugs inserted into the target.  The
composition of the sputtered material was determined to be
$12.0\pm 0.5\%$ by energy dispersive X-ray spectroscopy.  The
Curie temperature of thick films was measured to be 175 K, which
is consistent with this composition from earlier
measurements.\cite{KontosThesis}  The polycrystalline PdNi films
have the expected FCC crystal structure and grow with (111)
texture, as determined by X-ray diffraction from a 200-nm thick
film.  Similar multilayers of other FCC metals grown on thick Nb
base layers exhibit columnar growth, with columnar grains of width
$20-90$\,nm.\cite{Geng:1999}

\begin{figure}[htb]
\begin{center}
\includegraphics[width=3.25in]{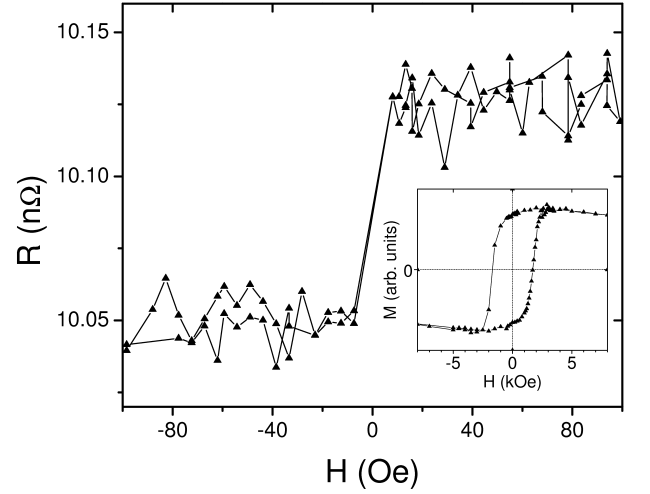}
\end{center}
\caption{$AR$ vs. $H$ data at 4.2 K for a Py/Cu/PdNi hybrid spin
valve with $d_{PdNi} = 12$\,nm.  Before taking these data, the
PdNi layer was magnetized in the negative direction by applying a
magnetic field of -5 kOe.  The Py layer switches when the applied
field is changed from -7.5 to +7.5 Oe.  The inset shows the
hysteresis loop at 10 K for a [PdNi(20)/Cu(10)]$_n$ multilayer
with $n=3$, where at saturation $M_{PdNi} = 116$\,emu/cm$^{^3}$.
The coercive field is about 1.5 kOe, much larger than the fields
used to switch the Py layer.} \label{Fig1}
\end{figure}

To measure $l_{sf}$ in the PdNi alloy, we sputtered hybrid spin
valves of the form Cu(10)/Py(24)/Cu(20)/PdNi($d_{PdNi}$)/Cu(10),
where all thickness are specified in nm.  In these samples PdNi
acts as the ``pinned" F layer and permalloy (Py =
Ni$_{84}$Fe$_{16}$) as the ``free" F layer. Because the coercive
field of PdNi ($H_c \approx 1.5 $\,kOe, see inset to Fig.
\ref{Fig1}) is much larger than that of Py ($H_c < 15$\,Oe), we
initially believed that exchange bias was not necessary to achieve
good antiparallel (AP) alignment of the PdNi and Py
magnetizations. (This issue will be discussed further below.) With
the current direction perpendicular to the planes (CPP), the
specific resistance is defined as $AR$ (sample cross-sectional
area times CPP resistance), and the specific magnetoresistance is
$A\Delta R = A(R^{AP}-R^{P})$.  The sample is sandwiched between
two $\sim$1-mm-wide superconducting Nb cross strips, giving $A
\sim$ 1 mm$^ 2$, and measured at 4.2\,K using SQUID electronics in
a current comparator bridge.  Our sputtering system and
measurement electronics are described in detail
elsewhere.\cite{Pratt:95}

A qualitative understanding of the magnetoresistance in our spin
valves can be gained from the simple two-current series-resistor
model.\cite{BassJMMM1999}  The model treats transport of up- and
down-spin electrons in parallel through the multilayer.  In the F
materials and at F/N interfaces, the majority and minority spins
generally have different resistivities and interface resistances,
defined below. When the thickness of the PdNi layer $d_{PdNi}$ is
much less than $l_{sf}^{PdNi}$, the two-current series-resistor
model\cite{BassJMMM1999} gives
\begin{equation}\label{thinlimit}
    A\Delta R \propto
    \beta_{PdNi}\rho^*_{PdNi}d_{PdNi}+\gamma_{PdNi/Cu}AR^*_{PdNi/Cu}.
\end{equation}
When $d_{PdNi} >> l_{sf}^{PdNi}$,
\begin{equation}\label{thicklimit}
    A\Delta R \propto
    \beta_{PdNi}\rho^*_{PdNi}l_{sf}^{PdNi}+\gamma_{PdNi/Cu}AR^*_{PdNi/Cu}.
\end{equation}
Thus, a plot of $A\Delta R$ vs. $d_{PdNi}$ will be linear at small
values of the PdNi thickness and saturate at large values.
Valet-Fert\cite{ValetFert:93} theory interpolates between these
two limiting cases, and the crossover between the two behaviors
can be used to extract $l_{sf}^{PdNi}$.  In Eqs. (\ref{thinlimit})
and (\ref{thicklimit}),
$\beta_{PdNi}=(\rho^{\downarrow}_{PdNi}-\rho^{\uparrow}_{PdNi})/(\rho^{\downarrow}_{PdNi}+\rho^{\uparrow}_{PdNi})$
is the bulk spin-scattering asymmetry, where
$\rho^{\uparrow(\downarrow)}_{PdNi}$ is the resistivity of the
PdNi for electrons with their moment parallel (antiparallel) to
the moment of the PdNi layer;
$\rho^*_{PdNi}=\rho_{PdNi}/(1-\beta_{PdNi}^2)$, and
$\rho_{PdNi}=(1/\rho^\uparrow_{PdNi}+1/\rho^\downarrow_{PdNi})^{-1}$
is the usual resistivity of a thick PdNi film. $AR^*_{PdNi/Cu}$
and the spin scattering asymmetry at the PdNi/Cu interface,
$\gamma_{PdNi/Cu}$, have similar definitions in terms of
$AR^\uparrow_{PdNi/Cu}$ and $AR^\downarrow_{PdNi/Cu}$, where
$AR_{PdNi/Cu}$ is the specific resistance of the PdNi/Cu
interface.

Fig. \ref{Fig1} shows the raw magnetoresistance signal for a
sample with $d_{PdNi}=12$\,nm.  The sample, immersed in liquid
helium, was first subjected to an applied field $H = -5$\,kOe to
align the PdNi magnetization ($M_{PdNi}$) in the negative
direction.  The field was then set to zero, and the sample was
slowly lifted in the cryostat to just above the liquid helium
level until a NbTi wire mounted near the sample transformed into
the normal state. Then the sample was lowered back into the liquid
helium.  This process removes any trapped magnetic flux from the
Nb contacts. Various tests indicate that $M_{PdNi}$ is not
affected by this warming. Once the sample was cold again, the
magnetic field was slowly swept from $H=-500$\,Oe to $H=500$\,Oe,
then back to $H=-500$\,Oe. The magnetization of the Py layer
switches abruptly when the field passes through zero, indicating
that its coercive field is less than the field step size of 7.5
Oe.  (The figure shows only the region between -100 and 100 Oe for
clarity.) The reproducibility of the data in the two sweep
directions indicates that $M_{PdNi}$ does not change significantly
during the sweep, due to the high coercivity of PdNi shown in the
inset.

\begin{figure}[htb]
\begin{center}
\includegraphics[width=2.8in]{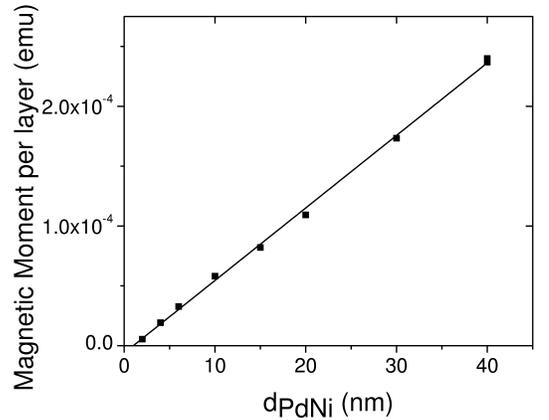}
\end{center}
\caption{Magnetic moment per layer of a series of PdNi/Cu
multilayers vs. PdNi thickness, at 10 K. The solid line is a best
fit straight line, whose x-axis intercept provides an estimate of
the total thickness of the magnetically dead layers at the two
PdNi/Cu interfaces.} \label{Fig2}
\end{figure}

An important issue in the magnetic behavior of thin films is the
possible existence of a magnetic ``dead" layer at the interface
with a nonmagnetic film, due to the reduced number of magnetic
atom nearest neighbors at the interface or to inter-diffusion or
alloying near the interface. To address this, we grew a series of
PdNi multilayers of the form
Nb(5)/Cu(10)/[PdNi($d_{PdNi}$)/Cu(10)]$_n$/Nb(5), where $d_{PdNi}$
varied between 2 and 40 nm, and $n$ was chosen to keep the total
PdNi thickness equal to 60 nm (except $n=1$ for
$d_{PdNi}=40$\,nm). The inset to Fig. \ref{Fig1} shows the
magnetization vs. field for the sample with $d_{PdNi}=20$\,nm,
measured at T = 10 K in a Quantum Design SQUID magnetometer, where
we have not subtracted out the small diamagnetic contribution from
the substrate, visible at high field.  Fig. \ref{Fig2} shows the
magnetic moment per layer (at saturation) vs. $d_{PdNi}$ for all
the multilayer samples. A linear fit extrapolates to zero at
$d_{PdNi}=1.1\pm0.4$\,nm, which we interpret as the total
thickness of the dead layers ($\delta_{dead}$) at the two PdNi/Cu
interfaces.

\begin{figure}[htb]
\begin{center}
\includegraphics[width=2.8in]{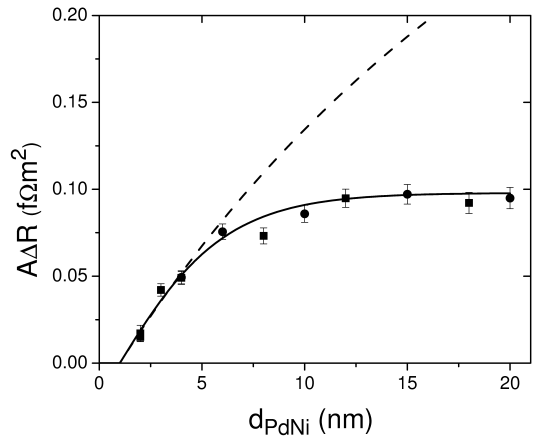}
\end{center}
\caption{$A\Delta R$ vs. $d_{PdNi}$.  The solid line represents a
fit to the data of the Valet-Fert equations, using fixed
parameters discussed in the text.  The fit provides an estimate of
the spin memory length in PdNi of $l_{sf}^{PdNi}=2.8\pm 0.5$\,nm.
The dashed line shows the result of a calculation assuming
$l_{sf}^{PdNi}=\infty$.} \label{Fig3}
\end{figure}

Fig. \ref{Fig3} shows the specific magnetoresistance, $A\Delta R$,
vs. PdNi thickness, for all of our samples.  The data start out
increasing nearly linearly with $d_{PdNi}$, then flatten out when
$d_{PdNi}$ exceeds about 10 nm.  The solid line is a fit to the
data of the Valet-Fert equations.\cite{ValetFert:93}  Many of the
material parameters used in the fit are obtained from previous
measurements.  These include the resistivity and bulk spin
scattering asymmetry for Py, $\rho_{Py}=123$\,n$\Omega$m and
$\beta_{Py}=0.76$,\cite{Vila:00} the specific resistance and spin
scattering asymmetry for the inner Py/Cu interface,
$AR^*_{Py/Cu}=0.50$\,f$\Omega$m$^2$ and $\gamma_{Py/Cu}=0.70$, and
the spin memory length in Py, $l_{sf}^{Py}=5.5$\,nm.\cite{Vila:00}
For the Cu layer between the Py and PdNi layers, we take
$\rho_{Cu}=4.5$\,n$\Omega$m\cite{Vila:00} and
$l_{sf}^{Cu}\geq500$\,nm.\cite{BassPrattReview} We estimate the
resistivity of PdNi alloy from Van der Pauw measurements on
isolated films of thickness 200 nm that give
$\rho_{PdNi}=121\pm$6\,n$\Omega$m. The interface resistance
between PdNi and Cu and the extent of spin memory loss at the
PdNi/Cu interface are estimated from earlier work on the Pd/Cu
interface,\cite{Galinon:05} which gave
$AR^*_{Pd/Cu}=0.42\pm0.07$\,f$\Omega$m$^2$ and
$\delta_{Pd/Cu}=0.24^{+0.06}_{-0.03}$, where the spin polarization
passing through the interface is reduced by the factor
$e^{-\delta}$. For simplicity, we assume no spin scattering
asymmetry at the inner PdNi/Cu interface, i.e.
$\gamma_{PdNi/Cu}=0$. The free parameters in the fit are then
$\beta_{PdNi}$, $AR_{PdNi/Cu/Nb}$,\cite{FNSnoasymm} and the
quantity we are most interested in, namely $l_{sf}^{PdNi}$, the
spin memory length in our PdNi alloy.

\begin{figure}[htb]
\begin{center}
\includegraphics[width=2.8in]{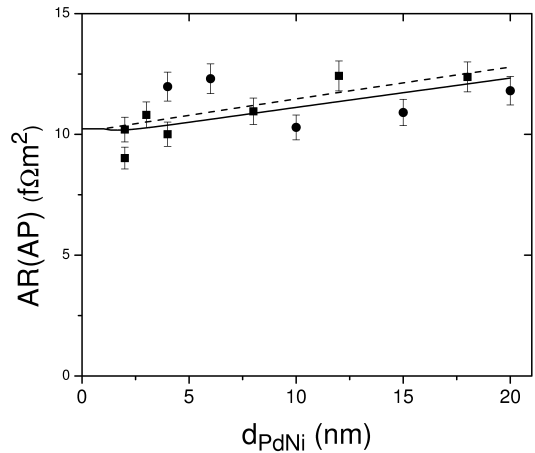}
\end{center}
\caption{$AR$ vs. $d_{PdNi}$ for the AP state.  The solid line is
the result of a calculation with the same parameters used to fit
the data in Fig. \ref{Fig3}.  The dashed line is for
$l_{sf}^{PdNi}=\infty$.} \label{Fig4}
\end{figure}

To further constrain the fitting parameters, we also fit the total
resistance of the spin valve in the antiparallel state, $AR^{AP}$,
shown in Fig. \ref{Fig4}.  For this, we need the interface
resistance between the Py and Nb layers (separated by the Cu
spacer, which is superconducting by proximity), known to be
$AR_{Py/Cu/Nb}=3.0$\,f$\Omega$m$^2$.\cite{Vila:00,FNSnoasymm}

In the numerical solution of the Valet-Fert equations, we include
a magnetically dead layer at the PdNi/Cu interfaces with total
thickness $\delta_{dead}$, following the discussion earlier.  The
best fit to the data, shown as the solid lines in Figs. \ref{Fig3}
and \ref{Fig4}, is obtained with $\delta_{dead}=1.0\pm0.2$\,nm, a
bulk spin-scattering asymmetry of $\beta_{PdNi}=0.14\pm0.02$, a
spin memory length in PdNi of $l_{sf}^{PdNi}=2.8\pm0.5$\,nm and a
PdNi/Nb interface resistance of
$AR_{PdNi/Cu/Nb}=2.3\pm0.3$\,f$\Omega$m$^2$.  These uncertainties
are based on the scatter in the data and the uncertainty in
$\rho_{PdNi}$. To illustrate the role of the finite spin memory
length in the fit, we show with the dashed line in Fig. \ref{Fig3}
the result of the calculation taking $l_{sf}^{PdNi}=\infty$.  The
value obtained for the PdNi/Nb interface resistance is consistent
with the value $AR_{PdNi/Nb}=2.31\pm0.07\,$f$\Omega$m$^2$ obtained
from the high-bias resistance of Josephson junctions of the form
Nb/PdNi/Nb, with PdNi thicknesses in the range 30-100
nm.\cite{Trupti}

In Fig. \ref{Fig4}, the dashed line, for $l_{sf}^{PdNi}=\infty$,
is not much different from the solid line for $l_{sf}^{PdNi}=2.8$.
Thus the fit to these $AR$ data is not very sensitive to the
choice of $l_{sf}^{PdNi}$.  The slope of the solid line agrees
reasonably well with data, indicating that our value of
$\rho_{PdNi}$ from van der Pauw measurements on 200-nm-thick films
is consistent with these CPP results. The value of
$AR_{PdNi/Cu/Nb}$ is determined primarily from this fit in Fig.
\ref{Fig4}, and its value has only a minor effect on the fit to
the $A\Delta R$ data of Fig. \ref{Fig3}.

The very short value of $l_{sf}^{PdNi}$ obtained from these
measurements was at first a surprise, because it is much shorter
than the decay length of the supercurrent in Nb/PdNi/Nb Josephson
junctions reported recently by some of us.\cite{Trupti}
Theoretical analysis of the Josephson junction data suggests that
the decay length should be equal to the mean free path in PdNi,
which was determined to be $l_e = 10.6 \pm 1$\,nm from a fit to
the critical current data.\cite{JJdecaylengths}  In the standard
picture of diffusive transport in metals, spin-flip and spin-orbit
scattering occur on length scales much longer than the mean free
path.  The data from these two experiments, however, indicate the
opposite behavior for our PdNi films, namely $l_{sf} < l_e$.  A
possible explanation for this unusual situation comes from our
recent discovery that the magnetic anisotropy in our thin PdNi
films is out-of-plane.\cite{Trupti} That discovery casts doubt on
our ability to produce a good AP state in our spin valve samples.
Indeed, the $M$ vs. $H$ data shown in the inset to Fig. \ref{Fig1}
are slightly rounded, with the remanent magnetization equal to
about 80\% of the saturation magnetization. We now know that this
rounding is due to the PdNi magnetization starting to rotate out
of the plane, because similar $M$ vs $H$ curves taken with $H$
pointing out of the plane are less rounded (see Fig. 3 of ref.
[\onlinecite{Trupti}]). The out-of-plane component of the PdNi
magnetization can cause rotation of spins initially aligned along
the in-plane Py magnetization direction, enhancing spin memory
loss in our spin valves.

To suppress the influence of the out-of-plane magnetic anisotropy
on our spin valve results, we have fabricated a second set of spin
valve samples of the form
Cu(10)/Py(24)/Cu(20)/PdNi($d_{PdNi}$)/Py(8)/FeMn(8) /Cu(10), with
$d_{PdNi}$ ranging from 0 to 20 nm. In these samples, the PdNi is
exchange coupled to the thin Py layer, which is in turn exchange
biased in-plane by the FeMn.\cite{NoguesJMMM1999} After
fabrication, these samples are heated to 170\,C (above the
blocking temperature of FeMn), then cooled to room temperature in
an in-plane field of 190\,Oe. Magnetization measurements on
PdNi/Py/FeMn test samples show that both the Py and PdNi are
exchange biased by this procedure. Thus the in-plane Py
magnetization constrains the PdNi magnetization to also stay
in-plane.  In these spin valves, the PdNi acts as a "spoiler"
layer\cite{Moreau:07} inserted into a high-performance Py/Cu/Py
spin valve, because $\beta_{PdNi} \ll \beta_{Py}$.  When the PdNi
layer is thin, the GMR signal is large, whereas as the PdNi layer
gets thicker, the signal diminishes until it finally reaches a
much smaller value corresponding to that of a Py/Cu/PdNi spin
valve. The Valet-Fert equations describe this process
quantitatively as $d_{PdNi}$ surpasses $l_{sf}^{PdNi}$.

\begin{figure}[htb]
\begin{center}
\includegraphics[width=3.25in]{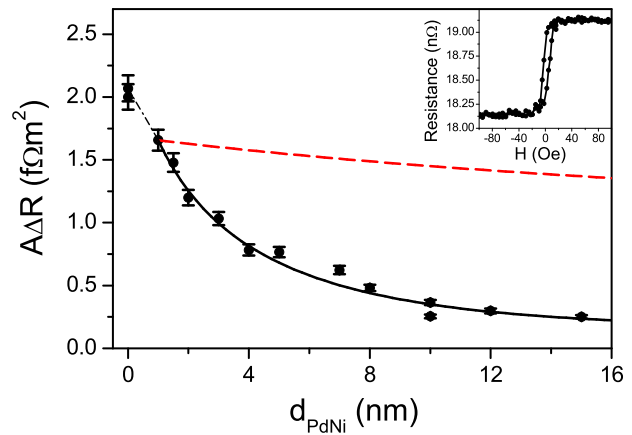}
\end{center}
\caption{$A\Delta R$ vs. $d_{PdNi}$ for "spoiler" spin valves of
the form Py/Cu/PdNi/Py/FeMn.  Without PdNi, these are
high-performance Py/Py spin valves with large $A\Delta R$. As the
PdNi thickness increases, $A\Delta R$ drops because of the low
spin-scattering asymmetry in PdNi relative to Py, i.e.
$\beta_{PdNi} \ll \beta_{Py}$. The black dot-dashed line shows a
linear interpolation between first two points. The black solid
line is a fit to the Valet-Fert equations for the rest of the
points, giving $l_{sf}^{PdNi} = 5.4\pm0.6$\,nm. The red
large-dashed line is obtained with $l_{sf}^{PdNi} = \infty$.
Inset: Raw magnetoresistance data for the sample with
$d_{PdNi}=2$\,nm.} \label{Spoiler}
\end{figure}

Fig. \ref{Spoiler} shows $A\Delta R$ vs. PdNi thickness for the
"spoiler" samples.  (Raw GMR data for the sample with
$d_{PdNi}=2$\,nm are shown in the inset.) As expected, $A\Delta R$
starts at a large value, and decreases with increasing PdNi
thickness. To extract $l_{sf}^{PdNi}$ from these data, we again
perform a numerical evaluation of the Valet-Fert equations.  The
data points at $d_{PdNi}=0$ are treated separately, since those
two samples (from separate sputtering runs) contain neither a
PdNi/Cu nor a PdNi/Py interface.  For the rest of the data points
starting at $d_{PdNi}=1.0$\,nm, we assume a magnetically 0.5-nm
thick dead layer at the PdNi/Cu interface following the earlier
discussion, and, for simplicity, a 0.5-nm-thick PdNi/Py interface.
We obtain a constraint between the interface parameters,
$\gamma_{PdNi/Py}$ and $AR^*_{PdNi/Py}$, by requiring the
calculation to agree with the $A\Delta R$ datum point at
$d_{PdNi}=1$\,nm. The overall $\chi^2$ is relatively insensitive
to the exact choice of $\gamma_{PdNi/Py}$ and $AR^*_{PdNi/Py}$
within the constraint; fortunately, however, the best-fit value of
$l_{sf}^{PdNi}$ changes little across the range of those interface
parameters. The fit leads us to the values $\gamma_{PdNi/Py}=
0.70\pm0.25$ and $AR^*_{PdNi/Py}=1.3\mp0.9$\,f$\Omega$m$^2$,
rather different from the Py/Pd interface parameters of
$\gamma_{Pd/Py}=0.41\pm0.14$ and
$AR^*_{Pd/Py}=0.2\pm0.1$\,f$\Omega$m$^2$.\cite{Sharma:07}  It
appears that the presence of Ni in the Pd significantly enhances
the PdNi/Py interface parameters, even though $\beta_{PdNi}$ is
very small.  Using the parameters already applied to the fit in
Fig. 3, with the exception of $l_{sf}^{PdNi}$, we obtain
$l_{sf}^{PdNi}=5.4\pm0.6$\,nm for this "spoiler" spin valve.

The longer value of $l_{sf}^{PdNi}$ obtained from the second set
of experiments supports the hypothesis that the PdNi magnetization
is highly non-uniform in the first set of spin valves with the
stand-alone PdNi fixed layer.  Electrons moving through the PdNi
experience an inhomogeneous exchange field, resulting in rapid
spin memory loss. This process may not be accompanied by efficient
momentum scattering, hence it allows the possibility of realizing
the regime $l_{sf} < l_e$. The fact that we obtain $l_{sf} < l_e$
also for the second set of samples seems to indicate the presence
of non-uniform magnetization even when the PdNi layer is
exchange-coupled to the adjacent Py layer with in-plane
magnetization, although the effect is smaller than in the first
set of samples.

What is the origin of the out-of-plane magnetic anisotropy in PdNi
films?  While carrying out these experiments, we learned of work
by Campbell and co-workers 30 years ago demonstrating that Ni
impurities in Pd exhibit a strong orbital magnetic moment, which
is indicative of strong spin-orbit
coupling.\cite{Campbell:1976,Campbell:1977} If the PdNi films grow
with uniform strain in the plane, the spin-orbit coupling may
induce out-of-plane magnetic anisotropy. (This depends on the
whether the strain is tensile or compressive in the plane, and the
sign of the spin-orbit coupling constant.) Furthermore, local
inhomogeneities in the strain due to crystalline defects or grain
boundaries could produce local fluctuations in the magnetic
anisotropy direction, which enhances spin memory loss as described
above. A surprising aspect of this story is the relatively long
decay length, 10.6 nm, of the critical current in Nb/PdNi/Nb
Josephson junctions.\cite{Trupti,JJdecaylengths} If our hypothesis
describing the local inhomogeneous magnetization in PdNi is
correct, then it appears that the Cooper pairs are not as
sensitive to magnetic disorder as is the single-electron spin
memory length. This is plausible given that the two spins in the
Cooper pair both precess in the same direction around the local
exchange field, maintaining their spin-singlet state in spite of
the rapidly changing exchange field.  Confirmation or refutation
of this picture will require a real theoretical calculation.

The measurements by Campbell and co-workers in the 1970's, in
addition to implicating strong spin-orbit coupling in PdNi alloy,
also showed that spin-orbit coupling in dilute PdFe alloys is
comparatively weak. In fact, a later study\cite{Williams:1992} of
the ac susceptibility of Pd$_{1-x}$Fe$_x$ alloys described the
alloy with x = 1.4\% as ``a nearly ideal system," in the sense
that the critical exponents at the Curie temperature follow
predictions of the Heisenberg model, and the magnetocrystalline
anisotropy is very small.  These observations motivated us to
measure $l_{sf}$ in PdFe, with the hope that it would be much
longer than in PdNi.

\begin{figure}[htb]
\begin{center}
\includegraphics[width=2.8in]{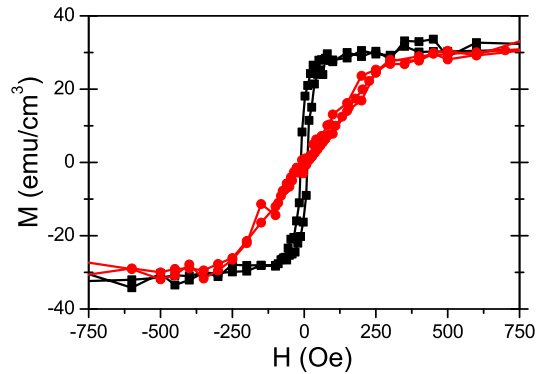}
\end{center}
\caption{(Color online) $M$ vs. $H$ at T=10 K for a 80 nm thick
PdFe film grown in a Nb(150)/Cu(10)/PdFe(80)/Cu(10)/Nb(25)
multilayer, with H pointing in-plane (black squares) and
out-of-plane (red circles).} \label{PdFeMvsH}
\end{figure}

\begin{figure}[htb]
\begin{center}
\includegraphics[width=2.8in]{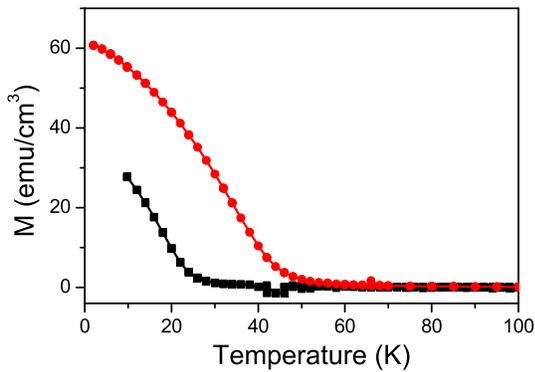}
\end{center}
\caption{(Color online) $M$ vs. $T$ for PdFe films of thickness 80
nm (black squares) and 400 nm (red circles). The films were
zero-field cooled to low temperature and then the magnetization
was measured by applying an in-plane field of 500 G. The
temperature was then increased gradually to 100 K keeping the
field constant. The Curie temperature of the 400 nm film indicated
an Fe concentration of about 1.1\%.} \label{PdFeMvsT}
\end{figure}

PdFe films were sputtered using a 5.6-cm diameter Pd target with a
single Fe plug of diameter 6.6 mm in the center. Our magnetization
measurements on PdFe films support the idea of low
magnetocrystalline anisotropy.  Fig. \ref{PdFeMvsH} shows $M$ vs.
$H$ data at $T=10$\,K for a 80 nm thick PdFe film grown in the
form Nb(150)/Cu(10)/PdFe(80)/Cu(10)/Nb(25), with $H$ both in-plane
and out-of-plane.  Relatively sharp switching to the saturation
magnetization is observed for $H$ in-plane, whereas for $H$
out-of-plane saturation isn't achieved until $H$ exceeds the
demagnetizing field of $\approx 300$Oe.  The strong preference for
in-plane magnetization indicates dominance of shape anisotropy
over magnetocrystalline anisotropy in these PdFe thin films.
Similar indications of in-plane anisotropy were also seen in a
60\,nm thick film of similar structure to the 80\,nm film and in a
stand-alone 400\,nm thick PdFe film (data not shown). The
temperature dependence of the magnetization for the 80 and 400\,nm
PdFe films is shown in Fig. \ref{PdFeMvsT}. The 80\,nm film showed
smaller magnetization and Curie temperature compared to the
400\,nm film, which could be due to strain during growth. The
Curie temperature of 44\,K obtained from the 400\,nm thick film
provides an estimate of the Fe concentration of 1.1 atomic
\%.\cite{Williams:1992}  A direct measurement of the Fe
concentration by energy dispersive X-ray spectroscopy on the same
sample yielded a somewhat larger result, 1.6 at. \%, but with a
large uncertainty due to the very small Fe concentration. The
in-plane saturation magnetization of PdFe measured on the 400\,nm
film at 10\,K was $M_{PdNi} = 63$\,emu/cm$^{^3}$=0.1 $\mu_B$/Fe
atom, which according to ref. \onlinecite{Williams:1992}
corresponds to an Fe concentration of approximately 1.4 at. \%.
These three determinations of the Fe concentration are reasonably
consistent, and lead us to our best estimate of 1.3 at. \%.

To measure $l_{sf}$ of PdFe, we sputtered hybrid spin valves of
the form Cu(5)/FeMn(8)/Py(8)/Cu(10)/PdFe($d_{PdFe}$)/Cu(10)/
Py(8)/FeMn(8)/Cu(5), again sandwiched between thick Nb electrodes.
The finished samples are heated to 170\,C and cooled in a field of
190\,Oe to pin the two Py layers to the FeMn via exchange bias.
These samples can be thought of as two FeMn/Py/Cu/PdFe hybrid spin
valves back-to-back, with their magnetoresistance signals adding.
The advantages of this geometry relative to the first geometry we
used with PdNi are that the magnetoresistance signal is twice as
large, and there are no interfaces between PdFe and Nb, hence
there are fewer parameters to take from previous work.

\begin{figure}[htb]
\begin{center}
\includegraphics[width=3.25in]{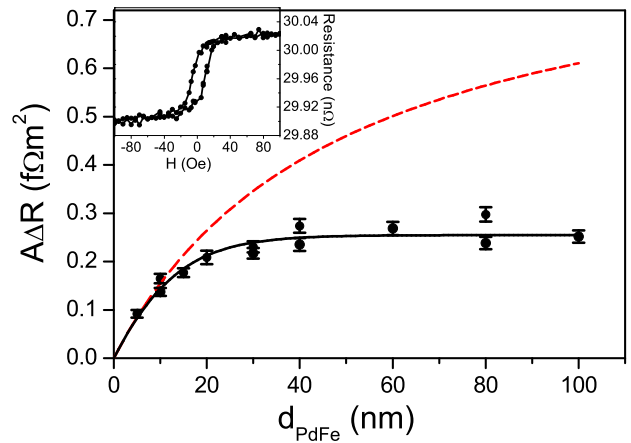}
\end{center}
\caption{$A\Delta R$ vs. $d_{PdFe}$ for the double hybrid spin
valves containing Py and PdFe alloy. The black solid line is a fit
to the data of the Valet-Fert equations, with the spin memory
length in PdFe of $l_{sf}^{PdFe}=9.6\pm 2.0$\,nm. The red dashed
line shows the result of a calculation assuming
$l_{sf}^{PdFe}=\infty$. Inset: Raw magnetoresistance data for the
sample with $d_{PdFe}=15$nm.} \label{PdFeFit}
\end{figure}

Raw magnetoresistance data are shown in the inset to Fig.
\ref{PdFeFit} for the sample with $d_{PdFe}=15 $nm.  As expected,
the PdFe layer switches its magnetization at a small field.  A
plot of $A\Delta R$ vs. $d_{PdFe}$ is shown in Fig. \ref{PdFeFit},
along with a fit to the Valet-Fert equations.  For this fit, the
value of $\rho_{PdFe}$ is needed.  From separate measurements on
100- and 200-nm-thick PdFe films, we have
$\rho_{PdFe}=85\pm4$\,n$\Omega$m. For the PdFe/Cu interfaces, we
use the same parameters as with PdNi/Cu. Fig. 8 shows the fit
assuming that $\delta_{dead}=0$ for the PdFe/Cu interfaces. If
$\delta_{dead}=1$\,nm, as assumed for the PdNi/Cu interfaces, the
fit in Fig. 8 is slightly changed at small $d_{PdFe}$. Including
the effect of these two choices of $\delta_{dead}$ on the fits, we
obtain $\beta_{PdFe}=0.11\pm0.02$ and $l_{sf}^{PdFe}=9.6\pm2$\,nm.

It is instructive to ask whether the values of the spin memory
lengths obtained in this work can be estimated from a previous
measurement of the spin memory length in nominally pure
Pd,\cite{Kurt:02} combined with the assumption that the spin
memory length is proportional to the mean free path, hence
inversely proportional to the resistivity.\cite{BassPrattReview}
The previous measurements on sputtered Pd gave
$\rho_{Pd}=42\pm2$\,n$\Omega$m and
$l_{sf}^{Pd}=25^{+10}_{-5}$\,nm.\cite{Kurt:02}  With these values,
we expect $l_{sf}^{PdFe} \approx
l_{sf}^{Pd}*(\rho_{Pd}/\rho_{PdFe}) \approx 12^{+5}_{-2.6}$\,nm,
and $l_{sf}^{PdNi} \approx l_{sf}^{Pd}*(\rho_{Pd}/\rho_{PdNi})
\approx 8.7^{+3.5}_{-1.8}$\,nm.  The former overlaps our measured
value of $l_{sf}^{PdFe}=9.6\pm2$\,nm, whereas the latter is still
a bit larger than our measured value of
$l_{sf}^{PdNi}=5.4\pm0.6$\,nm.  Perhaps this disagreement is due
to the stronger spin-orbit scattering in
PdNi.\cite{Campbell:1976,Campbell:1977} It is unfortunate that the
mean free path in Pd and its alloys is difficult to determine
independently.\cite{Trupti} Indeed, if it is true that
$l_{sf}^{PdNi} < l_e^{PdNi}$, as discussed earlier, then the whole
argument leading to the relationship above might be invalid.

In conclusion, we have measured the spin memory length in
Pd$_{0.88}$Ni$_{0.12}$ and Pd$_{0.987}$Fe$_{0.013}$ alloys using
giant magnetoresistance methods.  For PdNi, two different methods
were used, resulting in the values $l_{sf}^{PdNi}=2.8\pm 0.5$\,nm
and $5.4\pm 0.6$\,nm.  We believe that the smaller of these values
results from a conflict between the in-plane quantization axis
imposed by the experimental geometry and data analysis and the
recently-discovered out-of-plane magnetic anisotropy in PdNi
films.  This conflict produces inhomogeneous magnetization in the
samples, which in turn results in severe spin memory loss.  The
effect is mitigated somewhat when the PdNi magnetization is pinned
by an adjacent Py layer.  In PdFe, we find $l_{sf}^{PdFe}=9.6\pm
2$\,nm, a considerably larger value than in PdNi.

In closing, we note that the spin memory lengths measured in these
Pd alloys are considerably longer than what is found in another
popular weakly ferromagnetic alloy,
CuNi.\cite{Oboznov:06,Pratt:unpub}  If one wants to use one of
these alloys in a search for superconducting correlations with
spin-triplet symmetry,\cite{Bergeret:03} however, one must beware
of their limitations.  The cleanest signature of the so-called
``long-range triplet correlations" would be observation of a
proximity or Josephson effect penetrating a ferromagnet over a
distance much longer than the length scale associated with
conventional spin-singlet correlations.  The latter scale is
usually the larger of the mean free path, $l_e$, or the
dirty-limit exchange length, $\xi_F = \surd(\hbar D/E_{ex})$. As
discussed earlier in this paper, it appears that $l_{sf}^{PdNi}$
is \textit{shorter} than $l_e^{PdNi}$, eliminating any hope of
observing long-range triplet correlations in PdNi alloy.  In PdFe
the situation is slightly better, but still not ideal, as we
expect the mean free path of the 1.4\% Fe alloy to be longer than
that of the 12\% Ni alloy.

Acknowledgments: We thank I.A. Campbell for bringing references
[\onlinecite{Campbell:1976,Campbell:1977}] to our attention and
for helpful discussions, and B. Bi for performing X-ray scattering
measurements on our PdNi films. This work was supported by the
US-DOE under grant DE-FG02-06ER46341.


\begin{thebibliography}{99}
\bibitem{JJdecaylengths} The critical current of the Nb/PdNi/Nb
Josephson junctions studied in [\onlinecite{Trupti}] actually
decayed exponentially with a decay length of $7.7 \pm 0.5$ nm. The
theoretical formula to describe the decay, however, contains a sum
over Matsubara frequencies.  The result is that the primary decay
constant in the fit, which should be equal to the mean free path,
was determined to be $10.6 \pm 1$ nm.
\end{thebibliography}
\end{document}